\documentstyle[preprint,epsf,aps]{revtex}

\begin{document}
 
\title
 { Phase Diagram of High-T$_c$ Superconductors from a Field Theory Model}

\author{J. V. Alvarez $^1$, J. Gonz\'alez $^2$, F. Guinea $^3$ and 
M. A. H. Vozmediano $^1$ }
\address{
        $^1$Escuela Polit\'ecnica Superior. 
        Universidad Carlos III. 
        Butarque 15.
        Legan\'es. 28913 Madrid. Spain.\\
        $^2$Instituto de Estructura de la Materia. 
        Consejo Superior de Investigaciones Cient{\'\i}ficas. 
        Serrano 123, 28006 Madrid. Spain. \\
        $^3$Instituto de Ciencia de Materiales. 
        Consejo Superior de Investigaciones Cient{\'\i}ficas. 
        Cantoblanco. 28049 Madrid. Spain. 
        }
\date{\today}
\maketitle
\begin{abstract}

The renormalization group approach to correlated fermions
is used  to  determine
the phase diagram of the oxide cuprates modeled by the
$t-t'$ Hubbard model at the Van Hove filling. Spin--dependent
interactions give rise to instabilities corresponding to
ferromagnetic, antiferromagnetic and d-wave
superconducting phases. Antiferromagnetism and d-wave
superconductivity arise from the same interactions, and compete
in the same region of parameter space.
\end{abstract}

\section{Introduction}

Ten years after their discovery, the study  
of the high-T$_c$ superconductors\cite{htc} continues being a major 
puzzle for theoreticians. Despite the accumulation and accuracy of
experimental data now at hand, the theoretical situation has not 
improved much since the early days of high-T$_c$, and many of the models 
proposed then are still at work with very few new ideas available 
\cite{rev}. The problem is tough because the cuprates are 
a system of highly correlated electrons interacting at an intermediate to
strong coupling regime.

The paradigm of the metallic behavior, the Landau Fermi liquid theory, 
\cite{landau}
fails to describe the ``normal'' state above the critical temperature, 
and the BCS theory of superconductivity even in its strong coupling 
formulation can not account for the high temperatures reached by these 
compounds. It is clear that, even if phonons do play some role, 
we must look for a pairing attraction of a different nature -- electronic
or magnetic -- in the cuprates.
 
As it is known, all the copper-oxide high-T$_c$ materials come from an 
insulating antiferromagnetic ``father'' compound (a Mott insulator) which
becomes superconducting upon either electron or hole doping.  This  
metal-insulator transition inspired Anderson \cite{and1} to propose the 
two-dimensional Hubbard model close to half filling as
a starting point to model the correlations in the cuprates. The 
importance of antiferromagnetic fluctuations led to the early proposal 
of the t-J model \cite{zhang}. Most of the theoretical efforts
in the field are 
devoted to study the various extended Hubbard models with the available 
techniques numerical or perturbative \cite{hub}. 
The issue of whether or not the Hubbard model supports superconducting 
instabilities and at which range of temperature and doping is one of the 
most active areas of research on the field.

One of the most prominent features in the physics of the cuprates is the
many different energy scales where interesting phenomena occur. From the 
beginning it was clear  that the pair formation takes place at a 
different energy scale than the superconducting transition \cite{lee};
recent -- as well as early -- photoemission experiments on hole-doped  
materials have confirmed the  existence of a  pseudogap  in the
underdoped regime (below the doping at which the highest transition 
temperature occurs or optimal doping) at an energy scale much higher than
the  transition temperature. On the other hand we have the various 
magnetic scales. In order to reach the physics responsible for  a given 
phenomenon, we must be able to integrate away irrelevant degrees of 
freedom. 

A related feature is the existence of different kinds of 
coexisting and possibly competing instabilities within a certain range of
the parameters. In particular, one of the proposed pairing mechanism 
in the cuprates relays on the competing spin density-wave and 
superconducting 
instabilities, the pairing would be induced by an incipient instability
of the spin density-wave type. Besides, weak coupling approaches
to the Hubbard model have shown that it is more likely to develop a
spin-density-wave instability than superconductivity at
half-filling \cite{schulz,dzya,zanchi}. Inclusion of a next-to
nearest neighbors coupling, the so--called Hubbard t--t' model, enlarge
the possible instabilities of the system and opens the door
for d--wave pairing instabilities.

The renormalization group (RG) approach to interacting fermions 
proposed in \cite{shankar,pol} is an optimal framework to deal
with this problem.

In recent years great effort has been devoted to study the role
of Van Hove singularities (VHS) in two-dimensional electron
liquids \cite{hove,schulz,dzya,mark,tsuei,epl,npb,ioffe}.
Most part of the interest stems from the 
evidence, gathered from photoemission experiments, that the hole-doped 
copper oxide superconductors tend to develop very flat bands near the 
Fermi level \cite{photo,gofron}.
Near a Van Hove singularity the fermion density of states diverges so 
even very weak interactions can produce large effects.  
A Van Hove singularity is a saddle point in the dispersion relation
of the electron states $\varepsilon ( {\bf k} )$. 
In its vicinity, the
density of states diverges logarithmically in two dimensions, 
and shows cusps in three dimensions. 
The logarithmic divergence leads to a singular screening of the
interactions, in the same way as for the 1D Luttinger 
liquid \cite{lutt}.

Van Hove singularities have been largely ignored in the past mostly because
they do not arise in three dimensions where they can be integrated to give a
finite density of states. Their influence in two--dimensional systems 
was minimized on the following basis: under a theoretical point of view
it was argued that
i) they are isolated points in a Fermi line, hence a zero measure set;
ii) the shape of the Fermi surface is a relevant parameter in the RG 
sense that gets renormalized by the interaction hence fixing the Fermi 
surface at a VHS appeared as a fine-tunning condition; finally,
there was no anomalous behavior whose explanation could be helped by invoking
the existence of VHS. It has also been argued that disorder effects
would spoil the d--wave pairing predicted by the Van Hove model.

The physics of the cuprates has changed the above points in various 
respects. First of all, there are photoemission spectra showing the 
existence of very flat bands close to the Fermi level in most 
underdoped cuprates what suggests a Fermi surface very close to a VHS. 
Then, different approaches  including a RG study 
\cite{epl} have shown that the Fermi surface of the two dimensional 
Hubbard model has a tendency to be pinned 
near the VHS. It was shown in \cite{epl} that for open systems,
the renormalization of the chemical
potential is such that the VHS filling is an attractive fixed point
of the renormalization group. For a range of  initial dopings  close to the
singularity the renormalized system flows towards it.
This result obviates the major theoretical objection refering the fine 
tunning. 
Finally, it has been shown in a recent publication 
\cite{disorder}, that disorder effects may reduce but do not eliminate
the electronic pairing induced by the VHS's.

Even if the VHS's are not responsible for the normal state anomalies of
the cuprates, it is worth studying them as  their presence can 
substantially alter the behavior of any model. In the case of the 
Hubbard model on the square lattice, the existence of two independent  
VHS's also provides new scattering channels for the low-energy modes  
what reinforces the possibility of anisotropic pairing of electronic 
origin.

In addition to the phenomenological interest in condensed matter physics,
the study of the Hubbard model filled up to the level of the VHS, poses 
very interesting questions to the RG procedure applied to a
quantum statistical model
which would not arise in 
a standard renormalizable quantum field theory and that deepens our 
understanding of the RG physics.

In  previous works \cite{epl,npb} a superconducting instability was 
found  in a simplified model of VHS with two singularities and 
spin-independent 
interactions. In this paper we look for the instabilities of the
repulsive  $t-t'$
Hubbard model, filled up to the level of the Van Hove
singularity, following the RG program of Refs.
\cite{shankar,pol}. Preliminary results on this work can be found in
\cite{fases}.

The organization of the paper is as follows. First we introduce the model
as comes from the continuum limit of the one-band Hubbard model, identify
the VHS's and classify all possible couplings. Next we briefly review the
RG procedure as applied in condensed matter physics. In the next section 
we study the renormalization of the bare couplings for the case of a 
local interaction. Section 4 is 
devoted to the physical implications of this work. We study the response 
functions of the system and get the phase diagram that they lead to. 
Section 5 contains a summary of the results, discussions and future work.

\section{The model}

The RG properties of the Hubbard model at the Van Hove filling have been 
presented in \cite{epl,npb}. We will here review its most prominent 
features. As it is known, the Hubbard model was designed to reproduce
the Mott transition \cite{mott} found in some metals.
It is originally defined in a 
lattice by the hamiltonian:
\begin{equation}
H_{Hub} = - B \sum_{<ij>,\sigma}  c^+_{i,\sigma}c_{j,\sigma}
+U\sum_i n_{i,\uparrow} n_{i,\downarrow} \;\;\;,
\label{ham}
\end{equation}
where B is the band width without correlation, that is, in the absence of
U, U is the intra-atomic energy, $\;U=<e^2/r_{12}>\;\;,\;\;
c^+_{i,\sigma}\;$ creates an electron at site $i$ with spin $\sigma$,
and $n_i$ is the particle density at site $i$, 
$n_{i,\uparrow} = c_{i,\uparrow}^+c_{i,\uparrow}$. Hubbard showed that 
with this hamiltonian the spectrum of quasiparticles splits into two bands
which overlap when the lattice spacing $a$ is 
$\;a < a_0 =B/U = 1.15\;$ describing an antiferromagnetic metal while it
describes an antiferromagnetic  insulator for 
larger values of $a$. Nagaoka pointed out that the system would only
be insulating exactly at half filling when the number $n$ of electrons 
equals the number of  lattice sites being metallic for any other filling 
(doping). He also noticed that it would show ferromagnetic behavior in 
the limit of small values of $B/U$.

We shall here use the simple one-band Hubbard model which has proven
to be a good starting point for the description of the band properties
of most cuprate materials. In a tight-binding 
approximation, (\ref{ham}) is written as
$$ H=\sum_{{\bf k},\sigma} \varepsilon({\bf k})\; c^+_{{\bf k},\sigma}
c_{{\bf k},\sigma}
\;+\;U\sum_i n_{{\bf k},\uparrow} n_{{\bf k},\downarrow} \;\;\;. $$
When defined on a square lattice which is appropriate for the $Cu-O$ 
planes of the cuprates, and for nearest--neighbor hoping, the dispersion
relation $\varepsilon({\bf k})$ is
$$\varepsilon({\bf k}) = -2t \;[\cos(k_x a)+\cos(k_y a)]\;\;\;,$$
where $a$ is the lattice constant.
The Fermi surface at half filling has a diamond shape  with  
saddle points located at the four corners of the Brillouin zone
$(0,\pm\pi)\,,\,(\pm\pi,0)\;$. 
It shows perfect nesting \cite{shankar}, i.e. two parts of the
Fermi surface run parallel over an entire edge separated by a
common vector ${\bf Q}$. This nesting induces a peak in the
joint density of states $$J({\bf Q})=\int d^2k
N(\varepsilon_{{\bf k}}) N(\varepsilon_{{\bf k}+{\bf Q}})$$
even when the individual single particle density of states (DOS)
is featureless. In contrast, near a VHS, the DOS has already a
strong peak such that if ${\bf Q}$ joins two VHS's, a large peak
in J is assured. This will show up in the spin or charge
susceptibilities to be discussed later. 

The global nesting property is not observed 
in the photoemission experiments. 
The saddle points observed in \cite{photo}
can be incorporated into the metallic regime of the Hubbard model
by introducing a next-nearest neighbor interaction \cite{ttp}
which modifies the dispersion relation as
\begin{equation}
\varepsilon({\bf k}) = -2t\;[ \cos(k_x a)+\cos(k_y a)]
-2t'\cos(k_x a)\cos(k_y a)-\mu-2t'
\;\;\;,
\label{disp}
\end{equation}
where we have included the chemical potential $\mu+2t'\;$.
\begin{figure}
\epsfysize=9cm
\epsfbox{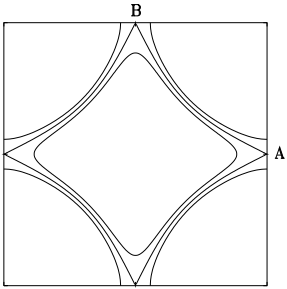}
\caption{Different shapes of the Fermi line for the $t-t'$
Hubbard model about the Van Hove filling.}
\label{contour}
\end{figure}

Different 
constant-energy lines for (\ref{disp}) are shown in
fig. 1. $\mu=0$
corresponds to the Fermi surface sitting at the
Van Hove singularities denoted A and B in the figure. 
The $t'$ interaction, besides
destroying the perfect nesting of the Hubbard model, allows to fit the 
phenomenology observed in the hole-doped cuprates for which the Fermi 
level lies  close to the saddle point at a doping of 0.15 to 0.25. 
$t'$ controls the shape and curvature of the Fermi surface. 
For the hole-doped materials, $t'<0$ and in all cases,
$t'<t/2$. The suppression of conventional nesting greatly
reduces the possibility of a charge density wave (CDW) or a spin
density wave (SDW) instability while retaining a strong
superconducting (SC) instability \cite{hirsch,ttp}

We construct a continuum model by assuming that for fillings such that 
the Fermi line lies close to the singularities, the majority of 
states participating in the interactions will come from regions in the 
vicinity of the points A, B of fig. 1. We then perform a Taylor 
expansion of (\ref{disp}) around A and B and shift the origin
of momenta to obtain the following 
dispersion relation
to be used as the kinetic term in the field theory model:
\begin{equation}
\varepsilon_{A,B} ( {\bf k} ) \approx \mp ( t \mp 2 t' ) k_x^2 a^2
\pm ( t \pm 2 t' ) k_y^2 a^2
\end{equation}
where  the momenta $k_x, k_y$ meassure
small deviations from $A, B$. As we see, the parameter $t'$
controls the angle between the 
two separatrices of the  hyperbolae of constant energy, 
$\varphi=2 \arctan  [ ( t + 2 t' ) / ( t - 2 t' ) ]$.

Knowing that scattering among the two singularities plays an
essential role in enhancing the superconducting instabilities 
of the system, we will map the problem
onto a model of electrons with two flavors A and B denoting the
electrons close to each of the VHS's. Interactions will take place among
electrons of the same and of different flavors.
We shall apply RG techniques to the model based on the fact that 
our physical system contains two natural energy scales. One 
is the bandwidth of the order of some electron volts and the other is the
temperature or the energy of the elementary excitations over the vacuum,
several orders of magnitude smaller. We will then establish an energy 
cutoff of the order of the band width and renormalize it towards the
Fermi surface $\varepsilon ({\bf k})=0$.

In the RG approach we write down a low-energy effective action which is
scale invariant at tree level and check where is it driven by the 
marginal and  relevant perturbations.  We will obtain RG
equations by integrating virtual states of two energy slices 
above and below the Fermi
level in an energy range given by the cutoff $E_c$,
$E_c - |d E_c| < |E| < E_c$. We are interested in the scaling
behavior of the interactions and correlations under a
progressive reduction of the cutoff, which leads to the
description of the low-energy physics about the Fermi level.

Integrating over a differential energy slice in the computation
of a given vertex function allows to extract the differential
equation that governs the scaling of the given coupling. The
procedure should be equivalent to the one used in
quantum field theory where the integration is performed over the
entire cutoff range and the derivative with respect to the
cutoff is taken afterwards. In our case, the differential approach
has the advantage that allows us to discuss scaling properties
without getting too close to the Fermi line ($E_c =0$) where
single--particle properties can become unreliable. We will insist
on that later in connection with the study of the response functions
of the system. 
 
Let us now proceed to the building of the model.
The free part of the low-energy effective action in momentum space is
\begin{equation}
S_0  =  \int d \omega d^2 k \sum_{\alpha,\sigma}\; [\omega
- \varepsilon_{\alpha} ( {\bf k})] \;
   a^{+}_{\alpha,\sigma}({\bf k}, \omega )
   a_{\alpha,\sigma}({\bf k}, \omega )  
\label{free}
\end{equation}
where $a_{\alpha,\sigma}\;\; (a^{+}_{\alpha,\sigma})$ is an electron
annihilation (creation) operator and $\alpha$ labels the Van Hove point.
The scaling behavior of (\ref{free}) has been analyzed in detail in 
\cite{npb}. It is clear that it is scale-invariant provided that 
under a rescaling of the energy 
$\;\omega\rightarrow s\omega\;\;,\;\; s<1 \;\;$,
we have
\begin{equation}
{\bf k}\rightarrow s^{1/2}{\bf k}\;\;\;,\;\;\;
a_{\alpha,\sigma}({\bf k}, \omega )\rightarrow s^{-3/2}\;
a_{\alpha,\sigma}({\bf k}, \omega )\;\;\;.
\label{sca}
\end{equation}
Next we discuss the interaction. 
In order to mimic a continuum analogue of the Hubbard interaction U 
in (\ref{ham}) we will write down in the effective action
an interaction of current--current type among currents of opposite
spin:
\begin{equation}
S_{{\rm int}}   =  - \frac{U}{2} \int d\omega d^2 k \; 
\left( \rho_{\uparrow } ({\bf k}, \omega)
 \: V( {\bf k}) \: 
      \rho_{\downarrow } (-{\bf k}, -\omega) \right)
\label{int}
\end{equation}
where $\rho_{\sigma }({\bf k}, \omega)$ are the Fourier
components of the density operator 
\begin{equation}
\rho_{\sigma} ({\bf k}, \omega) = 
 \frac{1}{(2 \pi)^3} \int d \omega_p d^2 p
\; a^{+}_{\sigma} ({\bf p} - {\bf k}, \omega_p - \omega )
   a_{\sigma} ( {\bf p},  \omega_p )
\end{equation}
In the spirit of the Wilson effective action,  U encodes all the 
possible couplings compatible with the symmetries of the problem
that are  marginal at tree level with the only extra requirement
of been among currents of opposite spin. We will not allow spin
flip interactions.
As mentioned before, the interaction parameter $U$ of the Hubbard model 
represents a point-like interaction of the fermions in the lattice. The 
direct interpretation of it in the continuum limit would be a 
delta--function interaction in real space, i.e. a constant 
$V( {\bf k})$ in Fourier space, or, else, we can interpret $U$ as a 
short--range interaction between the fermions having a finite support. 
It turns out that this two possibilities give rise to differences in the 
diagrammatics but do not alter the
physical realization of the model as we shall later see.

Here we will classify and analyze all spin-dependent local interactions
fixing $V( {\bf k}) = 1$. From the scaling behavior (\ref{sca}) we see 
that
for this  choice, the interaction is a marginal operator. Any analytical 
function $V({\bf k})$, once expanded in powers of ${\bf k}$, would leave
the constant term as a marginal operator and the rest would be irrelevant
operators. Notice that an interaction of the type discussed 
above, say
$V( {\bf r}) = 1/r^2$  could also arise as  the continuum limit of a 
local Hubbard interaction. In momentum space this is a logarithmic 
interaction of the kind that has been invoked before in the context of 
the Van Hove model \cite{npb} as a cure for the squared logarithms that 
appear in the renormalization of some diagrams. Such a logarithmic 
interaction is also scale invariant at tree level and should, in 
principle, be taken into account. The non--local character of the
$1/r^2$ interaction would exclude it from appearing in a quantum
field theory analysis where the locality of the operators is
linked to the property of causality. In a non--relativistic
model this consideration does not take place. We are then confronted
to study the renormalization of two different operators with the
same quantum numbers and the same scaling dimensions. We shall take
them as independent and will try not to mix them upon renormalization.
We will come to that later.

The RG analysis for the case of a 
spin-independent interaction with a finite support in $k$-space
and within a single singularity was done in detail in \cite{npb} while the
spin-independent case but allowing intersingularity scattering was
studied in \cite{epl}. The analysis presented here is the most complete
one performed within the RG and should show all the possible instabilities
of the repulsive t--t' Hubbard model for local interactions
at moderate values of U. 

At this point it is worth noticing that the marginal character of the 
four--fermion interaction in the two--dimensional Van Hove model
proposed here, marks already a difference with the isotropic
Fermi liquid model in two dimensions. There, the four--fermion
interaction is generically irrelevant being marginal only for
processes with a particular kinematics \cite{shankar}. 
This is due to the constraint that  momentum conservation imposes on the
different processes, very  severe in the case of
an isotropic Fermi line.
In particular
only forward and backward scattering gives rise to a --finite--
renormalization of the four--Fermi interaction in the standard
model. The reason is that these processes realize a kind of ``dimensional
reduction" and have the naive scaling dimensions of the D=1 situation.

In the Van Hove model, the four--Fermi coupling
is generically marginal due to the particular scaling of the
integration meassure (\ref{sca}) dictated by the free dispersion
relation. Although in the Fermi--liquid case the integration messure
also has the naive dimension of the D=1 case, there it is due to
the kinematical decomposition of the momenta into perpendicular
and parallel to the Fermi line and the choice that only the perpendicular
component scales with the energy. In the Van Hove case, the dimensional
reduction of the integration meassure has its origin in the 
particular form  of the Fermi line at the singularity very much
as happens in D=1.
 
Next, unlike what happens in the marginal couplings of the
Fermi liquid, the renormalization of the couplings in the Van
Hove model is nontrivial due to the logarithmic divergence  of the
density of states dictated again by the dispersion relation.
All that will become clear in what follows. 

The complete classification of the interactions including the two
flavors A and B follows exactly the one that occurs 
in the g-ology of one--dimensional systems \cite{lutt,libro} where the
role of the two Fermi points is here played by the two singularities.
We will see nevertheless that the one--dimensional parallelism stops
at the classification level as kinematical constraints in two dimensions
make the evolution of the couplings very different from that of 
one-dimensional systems. 
\vspace{0.5cm}

\begin{figure}
\epsfysize=12cm
\epsfbox{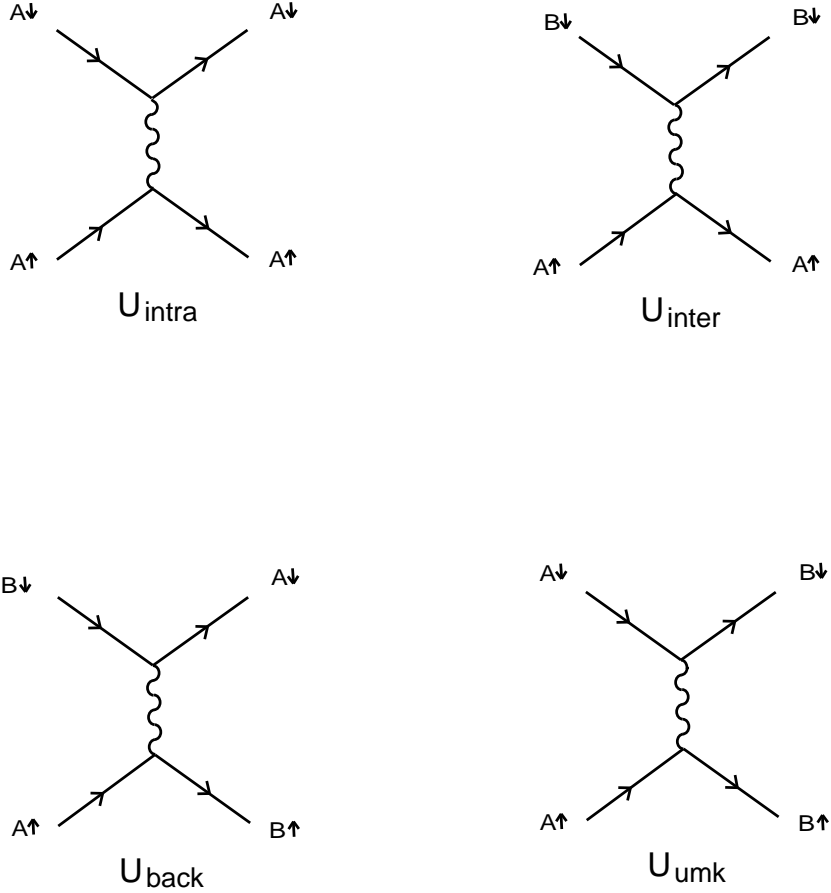}
\vspace{14pt}
\caption{Different interaction terms arising from the flavor indices 
A and B .}
\label{couplings}
\end{figure}

In general there are four types of interactions
that involve only low-energy modes. They are displayed in  
fig. 2  where 
the interaction is represented 
by a wavy line to clarify the process it refers to. In the model that we
are considering, the wavy lines should be shrank to a point giving rise
to couplings typical to the $\Phi^4$ quantum field theory.
The spin indices of the currents are understood to be opposite in
all cases. The interactions are classified as follows.
Intrasingularity interactions , $U_{{\rm intra}}$ occur around the 
same singularity. Low energy implies for that case low momentum transfer, 
the logarithmic singularities will occur at zero momentum transfer. 

Intersingularity interactions  occur when a type A current 
exchanges momenta with a type B current (or viceversa) as displayed in
fig. 2. The momenta exchanged in these processes is
again low.

A different kind of process occur when two electrons close to the 
singularities A, B, are excited to the vicinity of the opposite 
singularity. This is a low--energy process that involves a momentum 
transfer of order ${\bf Q}$, the vector joining the two singularities,
and the logarithmic singularities will appear for ${\bf Q} = (\pi,\pi)$.
The corresponding graph is called $U_{{\rm back}}$

The last interaction called $U_{{\rm umk}}$
deserves some comment. It describes a process  
in which two electrons of
opposite spin near the singularity $A$ jump together
to the singularity $B$. In the continuum, due to momentum 
conservation, such a process would not be allowed or else would
have a very  high energy. In the presence of a lattice
however the momentum needs only to be conserved modulo
one vector of the reciprocal lattice. In the $U_{{\rm umk}}$  interaction,
the momentum transfer ${\bf P} = (2\pi,2\pi)$ coincides with a lattice 
vector and is
such that $\varepsilon ({\bf k}+{\bf P})= -\varepsilon({\bf k})$.
Such interaction processes
are called umklapp. As we will see, they
play a major role in our model as they are responsible for the
antiferromagnetic and superconducting instabilities.

\section{Renormalization of the couplings.}

The interaction $V({\bf k})$ can be renormalized to second order
in perturbation theory by the diagrams despicted in 
fig. 3 where
the spin indices have been omited. The bare couplings of 
fig. 2 are by now
to be considered as vertex functions $i \Gamma ({\bf k}, \omega)$
i.e. the part of the interaction without the external legs.
\vspace{0.5cm}

\begin{figure}
\epsfysize=9cm
\epsfbox{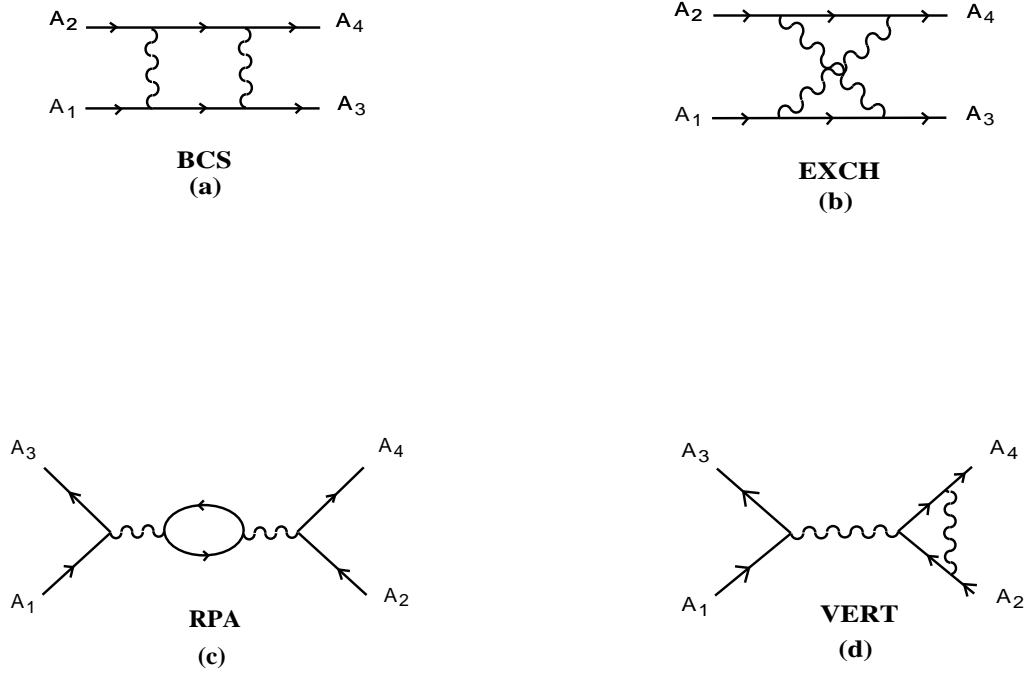}
\vspace{14 pt}
\caption{Diagrams contributing to the one-loop order correction 
to the interaction potential.}
\label{three}
\end{figure}
\vspace{0.1cm}
They are two types of corrections in fig. 3: direct (BCS)
and exchange (EXCH) particle--particle interactions 
(fig. 3 (a), (b)),
particle--hole interactions called RPA in fig. 3 (c), 
and vertex corrections called VERT in fig. 3 (d).

Once the interactions are shrinked to a point, they turn into the 
diagrams despicted in fig. 4 which are the ones to be computed 
in the one-loop calculation. We must keep in mind that, according to
fig. 3, the correction induced by the BCS diagram
(fig. 3 (c)) has a minus sign relative to the others as it
carries a closed fermion loop. 
\vspace{1cm}

\begin{figure}
\epsfysize=9cm
\epsfbox{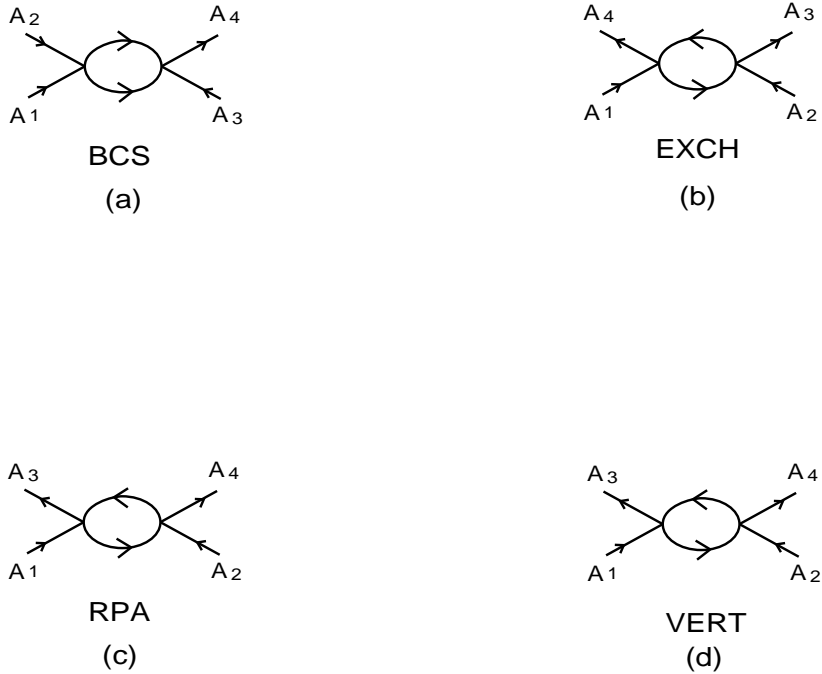}
\vspace{14 pt}
\caption{Diagrams of fig. 3 with the interaction
shrinked to a point.}
\label{four}
\end{figure}

The first thing to notice from fig. 4 is that
the particle--hole diagrams  (c), (d), when inserted 
into a coupling function, will provide exactly the same correction but 
with an opposite sign due to the fact that there is a closed fermionic  
loop in the original RPA coupling which is absent in the diagram 
called VERT. 
This cancellation that, to our knowledge, was first noticed in the
original work of ref. \cite{kl}, occurs to all orders in perturbation
theory. 
It has important consequences as it elliminates the RPA graphs
which are the typical screening processes for repulsive interactions
considered in most of the papers. It should be noticed however that the
cancellation only takes place for contact interactions of the Hubbard
type. Any k-dependence of the interaction  would restore the prevalence of
the RPA graphs for small momentum transfer. Moreover, diagrams (c) and (d)
of fig. 3 do not exist  if the interaction
is restricted  to currents of opposite spins as in our case.
In what this work is concerned, we are then restricted to study the 
vertex corrections provided by the particle--particle
diagrams of fig. 4 (a), (b).

The BCS diagram of fig. 4 (a) 
is only singular
for definite values of the external momenta and hence is not to be taken
into account in the renormalization of the vertex functions. Nevertheless,
it will
play an important role in the study of the instabilities of the
system through the response functions to be done in the next section.
The analysis of the renormalization induced by this coupling is similar 
to the one done for an isotropic Fermi line \cite{shankar}. The resulting
logarithmic singularity is of the form $\log\;({\bf k_1}+{\bf k_2})$ which
diverges only for the BCS kinematics, i.e. if the total incoming 
momentum adds to zero. In the differential approach that we are using,
this is best seen graphycally as despicted in fig. 5. 
\begin{figure}
\epsfysize=9cm
\epsfbox{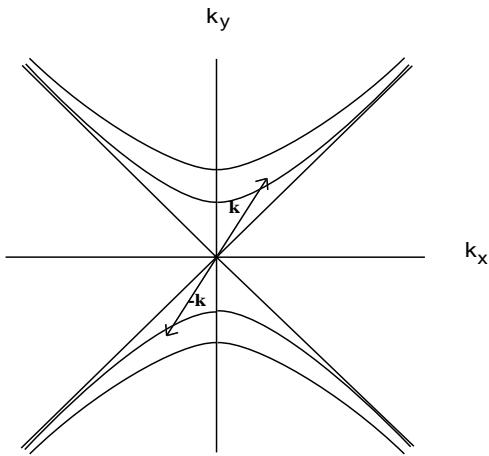}
\vspace{14pt}
\caption{Two energy slices used in the computation of the BCS graph.}
\label{slices}
\end{figure}

Two energy
integration  slices contributing to the computation of the
BCS graph are shown in fig. 5. It is clearly seeen that, unless
the total momentum of the incoming particles adds to zero, the
area of the intercept of the two bands, which meassures the cutoff
dependence of the diagram is of  order $(d E_c)^2$.
This is different from what happens in one dimension, where the BCS
graph contributes to the renormalization of all quartic couplings.
In the Van Hove model, 
the divergence of the BCS graphs at the kinematical singularity 
has a logarithm squared singularity but the kinematical
dependence remains the same.

This leaves us with the diagram of fig. 4
(b) as the only contribution to the renormalization of the
couplings at the one loop level. In terms of
the fermion propagator  $G^{(0)}_{\sigma} ({\bf k}, \omega )$ for each 
respective spin orientation, the vertex function at the one loop
level  $\: i \Gamma^{(2)} ({\bf k}, \omega) \:$ is
computed as
\begin{equation}
i \Gamma^{(2)}({\bf k},\omega) = - \frac{U^2}{(2 \pi)^3} 
\int_{-\infty}^{\infty} d\omega_q \int^{\Lambda} d^2 q \sum_{\sigma} 
G^{(0)}_{\sigma} ({\bf q} + {\bf k}, \omega_q + \omega )
G^{(0)}_{\sigma} ({\bf q}, \omega_q ) \;\;\;,
\label{vertex}
\end{equation}
where the momentum transfer in the vertices is such that 
$\Delta k_{2\rightarrow 3} = \Delta k_{1\rightarrow 4}$.
The momentum integrals are restricted to
modes within the energy cutoff, $|\varepsilon_{\alpha} \; ({\bf k})|
\leq E_c $ as the ones represented graphycally in fig. 5 where now
one of the slices is, as before, in the particle zone and the
second one is at the empty side of the Fermi sea. It can be
seen graphycally that the intercept of the two slices
for this case has always a linear contribution in $E_C$ for small
momentum transfer. 

The fermion propagator to be used in our model is
\begin{equation}
G^{(0)}_{\sigma} ({\bf q},\omega_q ) = \frac{1}{\omega_q -
\varepsilon({\bf q}) + i\epsilon \: \rm{sgn} \: \epsilon_q }\;\;\;.
\label{green}
\end{equation}
Near a Van Hove point, e.g. $\;A=(\pi,0)\;$, we have
\begin{equation}
\varepsilon ({\bf q}) = -u_0\;(q_x^2-\beta^2\;q_y^2)\;\;,
\label{disprel}
\end{equation}
where
$$u_0= 2 (t+2t')a^2 \;\;\;,\;\;\; \beta^2=\frac{t-2t'}{t+2t'} \;\;\;.$$
Perfect nesting will be absent as long as $\beta$ stays different from one
($t'\ne 0$).

Due to the sign of the imaginary part in (\ref{green}), 
the poles of the two propagators will be in a different half--plane only
if the particles involved come from opposite sign regions of 
$\varepsilon ({\bf q})$. This means that the virtual
states in the loop always involve a particle  (hole) on the filled 
$\varepsilon({\bf q})\;>0$  (empty)
side of the Fermi sea, been scattered to a hole 
$\varepsilon({\bf q}+{\bf k})\;<0$ 
(particle) on the opposite
region. 

Due to the presence of the two
flavors, two types of loops (polarizations) have to be computed: 
particle-hole processes around the same singularity and
processes in which the particle and the hole live near two different
singularities, the last involve a  momentum 
transfer of order ${\bf Q} = (\pi,\pi)\;$.

\begin{figure}
\epsfysize=4cm
\epsfbox{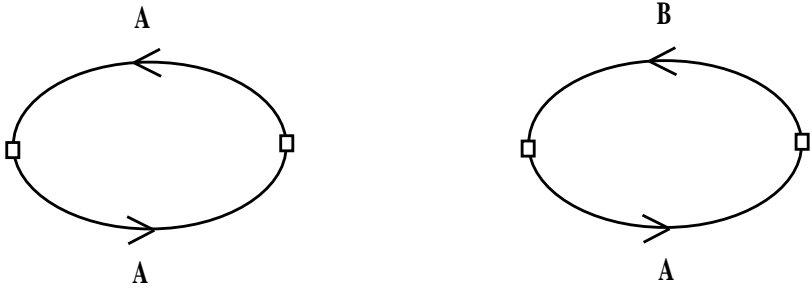}
\vspace{12pt}
\caption{The two polarizabilities used in the paper.}
\label{polar}
\end{figure}

The two polarizabilities shown in fig. 6
have been computed in \cite{epl,npb}. Their dependence on the cutoff is 

\begin{equation}
{\rm Re} \: \Gamma^{(2)}_{{\rm intra}} ( \omega) \sim 
\frac{c}{2 \pi^2 t} U^2  \log (E_c/\omega)
\label{re}
\end{equation}
\begin{equation}
{\rm Re} \: \Gamma^{(2)}_{{\rm inter}} (\omega) \sim
\frac{c'}{2 \pi^2 t} U^2 \log (E_c/\omega) \;\;,
\label{inter}
\end{equation}
where 
\begin{equation}
c \equiv 1/\sqrt{1 - 4(t'/t)^2} \;\;\;,\;\;\; c' \equiv
\log \left[ \left(1 + \sqrt{1 - 4(t'/t)^2} \right)/(2t'/t) \right]\;\;\;.
\label{ccp}
\end{equation}

As mentioned before, the logarithmic divergences of the vertex functions here
are due to the divergent density of states near the Van Hove singularity. 
The same graphs do not have a logarithmic dependence on the cutoff in
the case of the two--dimensional isotropic model for generic values of the
momenta. Only  forward or backward scattering at zero momentum are 
enhanced in that case \cite{landau}.
The bare susceptibility defined at this order in perturbation theory as
$\chi({\bf q})=
\Gamma({\bf q},\omega=0)$, diverges at both ${\bf q}=0$ where it
coincides with the density of states, and at ${\bf q}= (\pi,\pi)$,
due to intersingularity scattering. The latter has a squared
logarithm  singularity when the Fermi surface is nested,
$t'=0$, a situation that was treated in ref. \cite{npb}. This
squared logarithm singularity is cutoff to a usual logarithm 
when $t'\not= 0$.

The corrections to each of the couplings of fig. 2 are 
obtained by ``opening up'' the graph and inserting the polarizabilities
in such a way that the resulting graph is of the type of fig.
4 (b) and the vertices are made up of the tree--level 
interactions of fig. 2.

Let us first discuss the behavior of any coupling, say $U_{{\rm inter}}$.
$U_{{\rm inter}}$ is renormalized  by the diagrams shown
in fig. 7.

\begin{figure}
\epsfysize=16cm
\epsfbox{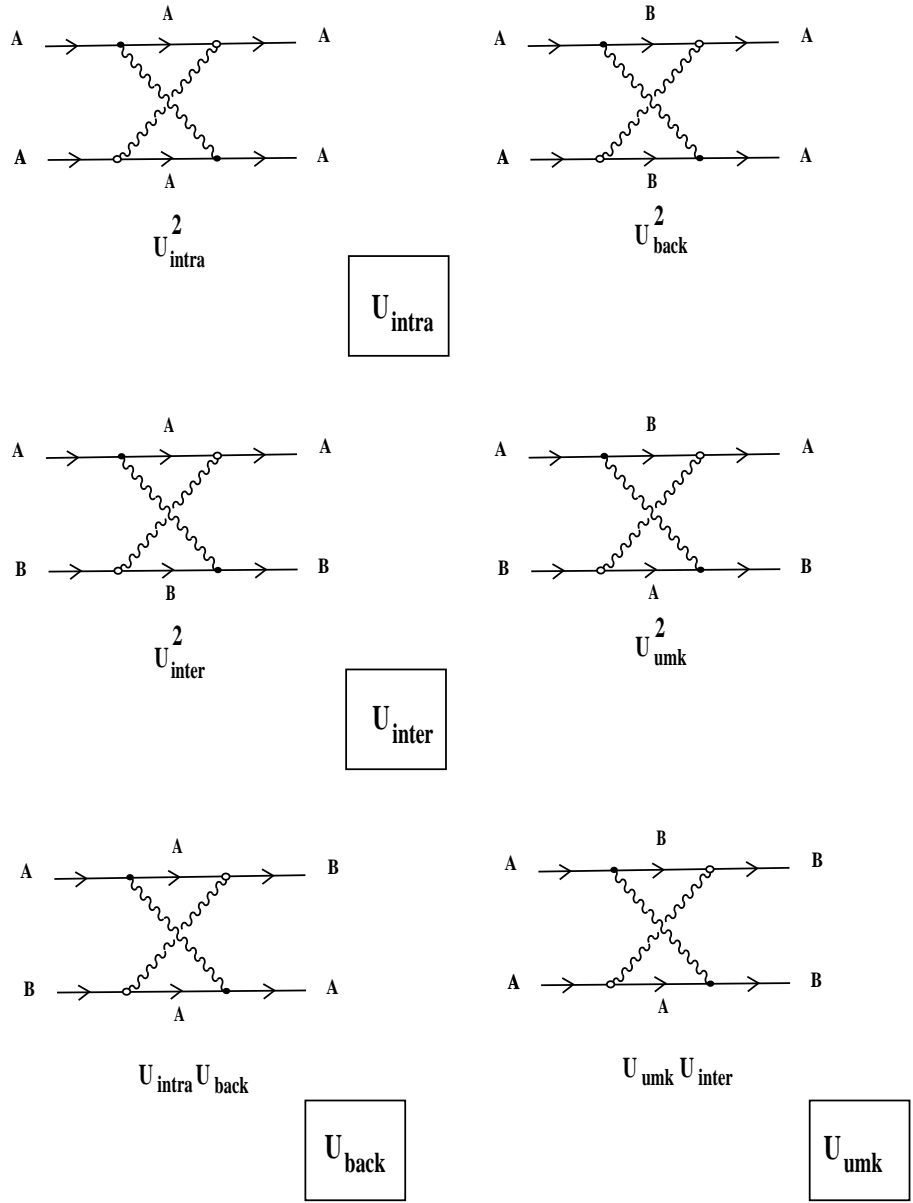}
\vspace{14pt}
\caption{Renormalization of the different couplings described in
the text.}
\label{corr}
\end{figure}

\vspace{0.5cm}
Adding up the one--loop correction to the bare coupling we find 
the vertex function at this order
\begin{equation}
\Gamma_{{\rm inter}}({\bf k},\omega) = U_{{\rm inter}}+
\frac{c'}{2 \pi^2t} ( U_{{\rm inter}}^2+U_{{\rm umk}}^2 )
\log \left|\frac{E_c}{\Lambda}\right|  \;\;\;E_c<\Lambda \;\;.
\end{equation}
Following the usual procedure \cite{amit}, we shall define the dressed
coupling constant at this level in such a way that the vertex function be 
cutoff independent what implies the RG equation
\begin{equation}
E_c\frac{d\; U_{{\rm inter}}(E_c)}{d \;E_c}=
\frac{c'}{2\pi^2t} ( U_{{\rm inter}}^2 + U_{{\rm umk}}^2 )\;,
\label{beta}
\end{equation}
were the polarizability involved is the interparticle polarizability
and the  sign of the beta function is positive as corresponds to the
``antiscreening'' diagram of fig. 3 (b). 
The same equation (\ref{beta}) is obtained by integrating
over a differential energy slice as we mentined earlier.

The growth of the coupling will eventually produce
an instability in the system to be discussed later.

We now apply the same procedure to the
rest of the couplings.
The diagrams that induce non--trivial renormalization 
are represented graphycally in fig. 7, where the spin 
polarizations of the currents have been omited and 
should be seen as  opposite in all cases.
We obtain the following set of coupled differential equations for
the couplings:
\begin{eqnarray}
E_c \frac{\partial U_{{\rm intra}}}{\partial E_c}  & = &
 \frac{1}{4\pi^2 t} c \left( U_{{\rm intra}}^2 + U_{{\rm back}}^{2}
       \right)    \label{flow1}       \\
E_c \frac{\partial U_{{\rm back}}}{\partial E_c}  & = &
 \frac{1}{2\pi^2 t} c \left( U_{{\rm intra}}  U_{{\rm back}} \right)
\label{flow2}     \\
E_c \frac{\partial U_{{\rm inter}}}{\partial E_c}  & = &
 \frac{1}{4\pi^2 t} c' \left( U_{{\rm inter}}^2 + U_{{\rm umk}}^{2} 
\right)\\
E_c \frac{\partial U_{{\rm umk}}}{\partial E_c}  & = &
 \frac{1}{2\pi^2 t} c' \left( U_{{\rm inter}} U_{{\rm umk}}
        \right)      \label{flow4}
\end{eqnarray}
where $c, c'$
are the prefactors of the polarizabilities at zero and ${\bf Q}$
momentum transfer, respectively given in (\ref{ccp}).

The RG equations
(\ref{flow1})-(\ref{flow4}) describe a flow that drives the
couplings to large values as the cutoff is sent to the Fermi
line. The growth of the couplings is to be understood as the tendency
of the system to flow towards a strongly coupled system
with different physical properties. Although the RG looses predictive
power as we approach the frequency
where the couplings diverge, the physical properties of this regime
can be qualitatively studied by means of the response functions
as will be described in the next chapter.

We will
compute the flow dictated by the RG equations
starting with all the couplings set to a common value
$U$. This assumption is not relevant in what concerns the
subsequent flow as long as the couplings start being positive. It
can be easily seen that the flow described by
(\ref{flow1})(\ref{flow2}) is attracted towards a region in which
$U_{intra}\sim U_{back}$ and that both diverge at the same
critical scale. The same applies to $U_{inter}$ and $U_{umk}$.
The only relevant feature may be a significant difference
between $U_{intra}$ and $U_{inter}$ at the starting point of the
RG flow. Under the change of variables

$$U_{1\pm}=U_{intra}\pm U_{back} \;\;\;,\;\;\; 
U_{2\pm}=U_{inter}\pm U_{umk} \;\;\;,$$
(\ref{flow1})--(\ref{flow4}) turn into
$$\frac{\partial U_{1\pm}}{\partial \log E_C}=\frac{c}{2\pi^2t}
U_{1\pm}^2 \;\;\;;\;\;\;
\frac{\partial U_{2\pm}}{\partial \log E_C}=\frac{c'}{2\pi^2t}
U_{2\pm}^2\;\;\;,$$
what shows that the above initial difference may be
reinterpreted in terms of an equivalent model with
 $U_{intra}=U_{inter}$ and different values  
of the constants c and c'.

Under the initial condition that all couplings are equal to $U$
it is clear that $U_{{\rm intra}} = U_{{\rm back}}$ 
and $U_{{\rm inter}} = U_{{\rm umk}}$, all along the flow. 

A scaling analysis of the equations (\ref{flow1})--(\ref{flow4}) trading the
cutoff dependence by a dependence on the momentum at which we are 
probing the system, allows us to write down the following equations

\begin{eqnarray}
U_{{\rm intra}}(\omega)  & = & \frac{U}{1 + U \: c/(2 \pi^2 t) \:  
\log(\omega
       /E_c) }     \label{intras}       \\
U_{{\rm inter}}(\omega)  & = & \frac{U}{1 + U \: c'/(2 \pi^2 t) \:  
\log(\omega
       /E_c) }      \label{inters}
\end{eqnarray}

The flow of the two couplings is despicted in fig. 8. 

\vspace{0.5cm}

\begin{figure}
\epsfysize=10cm
\epsfbox{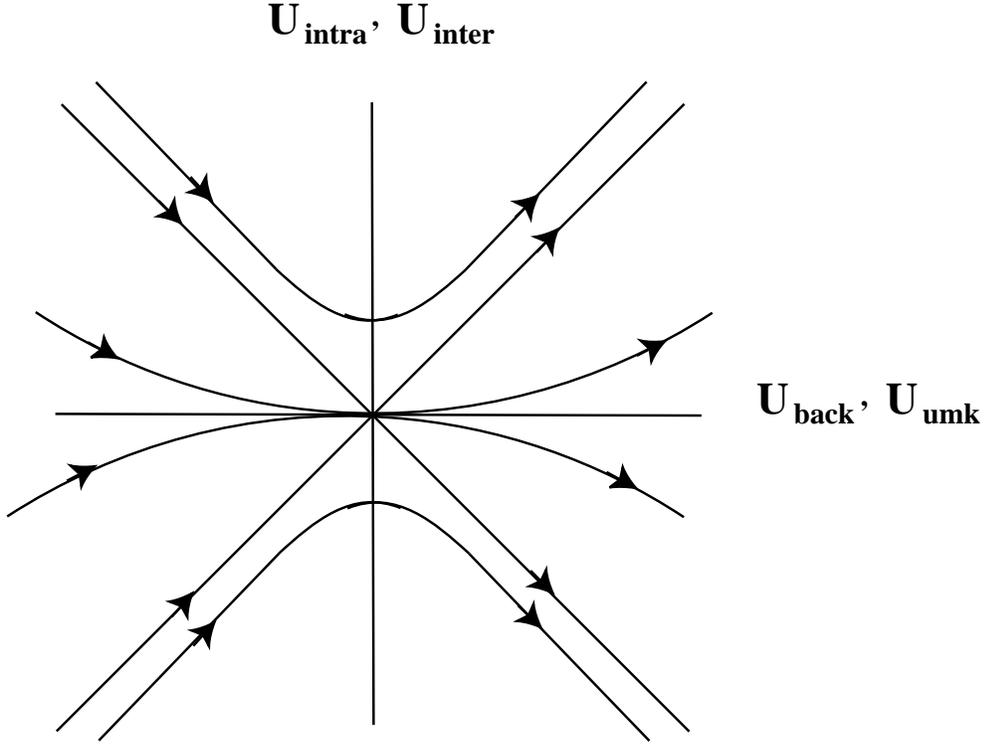}
\vspace{14 pt}
\caption{Schematic view of the flow of the 
couplings discussed in the text.}
\label{flow}
\end{figure}
\vspace{0.5cm}

We shall end this section with the discussion of
a fine point that arises
in the renormalization of the couplings at the one loop level.
In the study of the one--dimensional case, most references 
make a distinction between couplings among currents with different
spin orientation as the one considered here, called $U_{\perp}$,
and currents with the same spin orientation, called $U_{\parallel}$.
We have neglected the last set because we made the decision of
discussing a continuum model as  close as possible to the original
Hubbard model (\ref{ham}). We could have, in principle, enlarged
the model with the inclusion of 
two extra couplings of the type $U_{{\rm inter}\parallel}$
and  $U_{{\rm back}\parallel}$. The other two parallel couplings
 $U_{{\rm intra}\parallel}$ and  $U_{{\rm umk}\parallel}$, would be
forbidden at tree level by the Pauli exclusion principle and the
point--like character of the interaction.

\vspace{0.5cm}

\begin{figure}
\epsfysize=4cm
\epsfbox{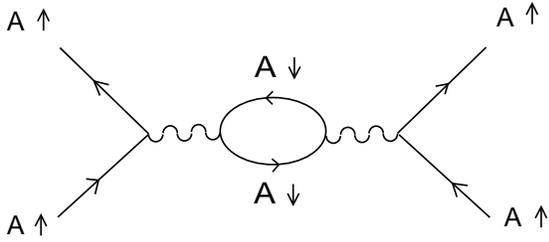}
\vspace{14 pt}
\caption{A diagram which generates a parallel coupling at one loop
level.}
\label{paralelo}
\end{figure}
\vspace{0.5cm}

The point is that by joining up the external legs of two
couplings, say of the $U_{{\rm intra}\perp}$ type as shown in
fig. 9 , 
we can generate
at one loop a parallel coupling of the type  $U_{{\rm intra}\parallel}$
not present in the tree level lagrangian. Generation of new  couplings
at higher orders in perturbation theory is usually protected in
quantum field theory (QFT) by symmetry principles --- unless the theory
is non--renormalizable --- but it is a well known phenomenon in
condensed matter systems responsible for physically relevant 
effects such as the Kondo effect or the attractive coupling among
electrons induced by the electron--phonon interaction. The usual
treatment in QFT would be to add to the lagrangian at tree level
the coupling whose renormalization is being established at one
loop. We can not do that in this case for the reasons mentioned
before, and we must interpret the whole phenomena as the fact that,
in the process of renormalization, extended interactions  of the type 
$1/r^2$ are generated, i.e. 
the point--like character of the interaction can not be
maintained in the renormalized theory. To construct a model fully
consistent we must then include momentum--dependent couplings 
which will eventually mix up in the renormalization of the local
interaction.

A detailed study of these new couplings will be the object of a
subsequent paper \cite{new}. To make plausible the phase diagram
that we obtain in this paper, we mention the fact that
the new couplings being momentum--dependent, they will be mostly
renormalized by the screening diagram of fig. 3c   which drives them
to zero if they start being repulsive as was shown in 
\cite{epl}. Their influence on the phase diagram will be 
discussed in the next section.

\section{The RG phases of the system.}

We now turn to the question of the phenomenological consequences of the
RG flow of fig. 7.
We interpret the divergences of the vertices in the same way as
in the RPA, as signalling the development of an
ordered phase in the system.
The precise determination of the instability which dominates for
given values of $U$ and $t'$ is accomplished by analyzing the
response functions of the system. The procedure
is similar to that followed in the study of one-dimensional
electron systems\cite{lutt,libro}.
 
The nature of the ground state of the
system is studied by means of the linear response functions or 
generalized susceptibilities. They describe the response of the system
to an external perturbation; a singularity in the response is an
indication that spontaneous distorsion or ordering can occur in the
system.  They are defined as the vacuum expectation value of the
correlation function of the given operator. A non-zero value 
signals the spontaneous breakdown of the symmetry associated to the
corresponding operator. 

The response function related to a given charge, spin
or superconductivity pairing operator ${\cal O}$ is defined by 
\begin{equation}
R(\omega, {\bf k})=-i\int dt e^{i\omega t}<{\cal O}(t, {\bf k}) 
{\cal O}^+(0, {\bf k})> \;.
\end{equation}
The operators of interest in our case will be
the following:
$${\cal O}_{CDW}(t, {\bf k}+{\bf Q})= \sum_{{\bf p},s}
\left[ a^+_{p+k+Q,s}(t)b_{p,s}(t) + h.c.\right] \;\;,$$
\begin{equation}
{\cal O}_{AFM}(t, {\bf k}+{\bf Q})= \sum_{\bf p}
\left[ a^+_{p+k+Q\uparrow}(t)
b_{p\uparrow}(t) - a^+_{p+k+Q\downarrow}(t)b_{p\downarrow}(t)
\right]\;\;
\end{equation}
$$\;\;\;\;= \rho_{\uparrow} ({\bf Q}, t ) -
\rho_{\downarrow} ({\bf Q}, t ) \;,$$
$${\cal O}_{FM}=
\rho_{\uparrow} (0, t ) -
\rho_{\downarrow} (0, t ) \;,$$
$${\cal O}_{SCS}(t,{\bf k})=\sum_{{\bf p}} \left[
a_{p+k \uparrow }(t) a_{-p \downarrow }(t) +
b_{p+k \uparrow }(t) b_{-p \downarrow }(t) +
 h.c. \right]\;,$$
$${\cal O}_{SCd}(t, {\bf k}) =\sum_{{\bf k}} \left[
a_{p+k \uparrow }(t) a_{-p \downarrow }(t) -
b_{p+k \uparrow }(t) b_{-p \downarrow }(t) +
 h.c. \right]\;,$$
where $a^+, b^+$ are creation operators for the particles at
A, B respectively.

${\cal O}_{CDW}$ is the order parameter associated to the formation
of a charge density wave instability with the wave vector ${\bf Q}$ joining
the two VH singularities. ${\cal O}_{AFM}$ is associated with the
formation of a spin density wave inducing an antiferromagnetic order
in the system. It will also expected to be singular at the value
${\bf Q}$ of the external momentum.  The operators
${\cal O}_{SCS}$, ${\cal O}_{SCd}$ are the order parameters associated
to singlet pairing with S and d--wave symmetry respectively; 
${\cal O}_{FM}$ describes ferromagnetic order. These later operators
are expected to be singular at zero momentum.

We will study  the response functions using the same  procedure
as described in \cite{schulz} for D=1.
Let us choose as an example the antiferromagnetic
response function ${\cal R}_{AFM}$. 
At a given value of the cutoff, the perturbation series for the
the AFM response function ${\cal R}$ has the general structure
represented graphycally in fig. 10.

\vspace{0.5cm}

\begin{figure}
\epsfysize=9cm
\epsfbox{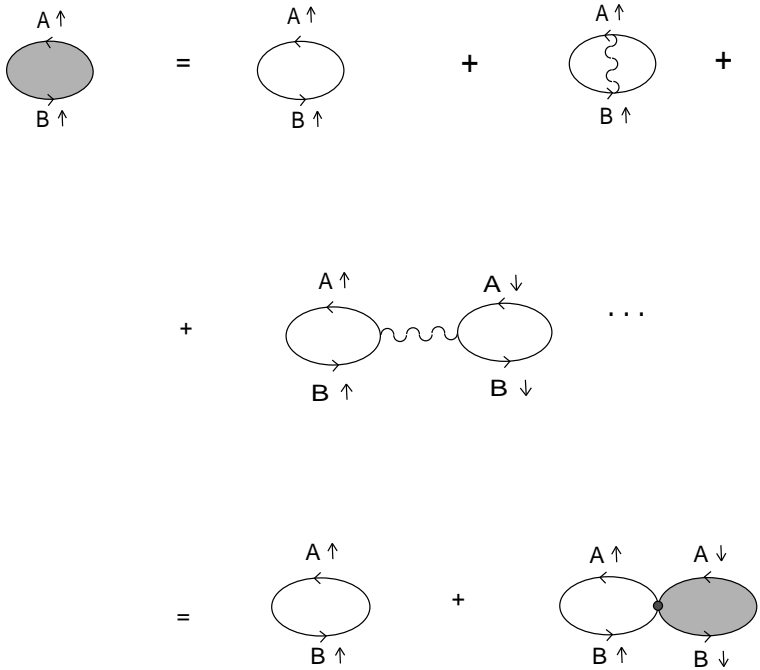}
\vspace{14 pt}
\caption{Perturbation series of the AFM response function 
truncated at order U.}
\label{response}
\end{figure}
\vspace{0.5cm}

In our case, due to the absence of parallel couplings, the
second class of diagrams in the first line of fig. 10 do not
participate in the first order correction.
The first perturbative term for ${\cal R}_{AFM}$
is built from a couple of one-loop
particle-hole diagrams linked by the interaction.
Each particle-hole
bubble has a logarithmic dependence on the cutoff $E_c$, with
the prefactor c' of (\ref{inter}).
The iteration
of bubbles can be taken into account by differentiating with
respect to $E_c$ and writing a self-consistent equation for
${\cal R}$ where the couplings and ${\cal R}$ are to be taken
as the renormalized values obtained after integrating out a
stripe of high--energy modes. 
In the case of the AFM response function we obtain the equation
\begin{equation}
\frac{\partial R_{AFM}}{\partial E_c} = -  \frac{c'}{\pi^2 t}
  \frac{1}{E_c}  +  \frac{c'}{2 \pi^2 t} \left( U_{back} +
  U_{umk} \right) \frac{1}{E_c} R_{AFM} \;\;,
\label{aferro}
\end{equation}
The couplings involved in this case are shown in fig. 11 to be
$\; U_{AFM}=U_{back}+U_{umk}$. 
\begin{figure}
\epsfysize=5cm
\epsfbox{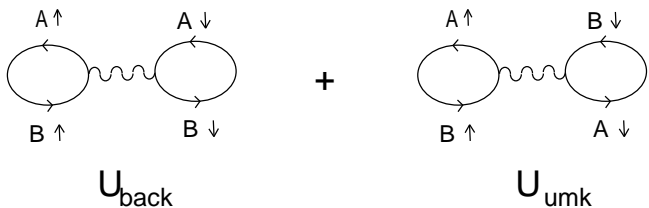}
\vspace{14 pt}
\caption{The effective coupling constant involved in the
AFM response function.}
\label{cafm}
\end{figure}
\vspace{0.5cm}

These couplings are the renormalized couplings
given by (\ref{intras}), (\ref{inters}). In the derivation of (\ref{aferro})
we have substituted the value of ${\cal R}_0$ by its renormalized
value ${\cal R}$.

At this stage it is worth to notice the difference of this procedure
with a usual RPA computation. In view of fig. 10, we could be
tempted to write down an integral equation for the complete 
response function of the type
$${\cal R} = {\cal R}_0 - U{\cal R}_0{\cal R} = {\cal
R}_0(1-U{\cal R}) \;\;,$$
which can be summed up to give the RPA result. This procedure relies
strongly upon the knowledge of the response function in the absence
of interaction and on the assumption that the interacting quantities
can be obtained adiabatically from the non--interacting ones. This
is correct in the case of Fermi--liquid like systems where the
single--particle properties of the full system are connected adiabatically
with the vacuum of the free system. Such a procedure breaks down
completely in the D=1 case where the interaction renormalizes strongly
the single--particle properties. The Van Hove system that we are
studying resembles the D=1 case in the respect that quasiparticles
cease to make sense once we get too close to the Fermi line,
i.e, when the cutoff is taken to zero. The particle--hole buble
computed in this case does not make sense and can not be used as
a starting point of a perturbative computation. What we do
instead, similar to what is done in the D=1 case, is to rely
upon the values of the naked response function for an energy close
to the value of the cutoff $E_c$ 
where the Fermi--liquid behavior of the system
can be assumed. We compute the scaling evolution of the response
function upon integration of a strip of high--energy modes
well above the Fermi line where the free Fermi propagator does
make sense and obtain a differential equation for the response
functions which can describe the physical properties of the system
for energies below the critical frequencies where the couplings 
diverge.

The study of the ferromagnetic response function can be done
following the same steps. It is given by 
the correlation of
the uniform magnetization, $\rho_{\uparrow} (0,\omega ) -
\rho_{\downarrow} (0,\omega ) $.
The particle-hole
bubble in this case has
the prefactor $c = 1/\sqrt{1 - 4(t'/t)^2}$. The equation for
$R_{FM}$ is
\begin{equation}
\frac{\partial R_{FM}}{\partial E_c} = -  \frac{c}{\pi^2 t}
  \frac{1}{E_c}  +  \frac{c}{2 \pi^2 t} \left( U_{intras} +
  U_{inters} \right) \frac{1}{E_c} R_{FM} \;\;\;.
\label{ferro}
\end{equation}

It is easily seen that the operators related to
charge-density-wave and s-wave superconducting instabilities do
not develop divergent correlations at small $\omega $. 
The scaling equation for the response function ${\cal R}_{CDW}$
computed with the previous techniques is 
\begin{equation}
\frac{\partial R_{CDW}}{\partial E_c} = -  \frac{c'}{\pi^2 t}
  \frac{1}{E_c}  -  \frac{c'}{2 \pi^2 t} \left( U_{back} +
  U_{umk} \right) \frac{1}{E_c} R_{CDW} \;\;\;,
\end{equation}
and goes to zero as the couplings grow large. The same happens
with ${\cal R}_{SCS}$.

The phase
diagram in the $t'-U$ plane is drawn by looking at the
competition among ferromagnetic, antiferromagnetic and d-wave
superconducting instabilities.

We recall that $U_{intras} +
U_{inters}$ and $U_{back} + U_{umk}$ have the same flow, within
the present model. Therefore, we may discern at once that
whenever $c > c'$ the ferromagnetic response function $R_{FM}$
prevails over $R_{AFM}$.

The study of the d--wave superconducting response function differs
from the previous ones in various respects. First,
as can be seen from the
definition of ${\cal R}_{SCD}$, it is the second  kind of diagrams
in the first line of 
fig. 10 what contributes to the renormalization of the response
function at first order in U. This implies
that the effective coupling constant in this case has a
relative minus sign being $U_{SCd}=U_{umk}-U_{intra}$.
Next and more important,  
the dependence on the cutoff of the
correlator $R^0_{SCD}$ given by
the diagrams at strictly zero total momentum has a log squared singularity
of the form
$\log^2 (E_c/\omega)$. We can still derive scaling  equations in
this case due to the fact that the  dependence of the equation on the
energy and the cutoff maintains the same same functional form what allows us 
to trade the cutoff by the energy.

The  equation for $R_{SCd}$ reads then
\begin{equation}
\frac{\partial R_{SCd}}{\partial E_c} = -  \frac{c}{2\pi^2 t}
  \frac{\log (E_c/\omega) }{E_c}  -
  \frac{c}{2 \pi^2 t} \left( U_{intra} -
  U_{umk} \right) \frac{\log (E_c/\omega) }{E_c} R_{SCd}
\label{scd}
\end{equation}
This equation also shows a homogeneous scaling of $R_{SCd}$ on
$\omega / E_c$, like in the previous cases.

We shall interpret the previous equations in the sense that, for
a given t' and U, the system will be in a phase given by the 
response function that divergest first -- or grows larger --
at the largest  frequency. The frequencies can be obtained by
integrating  the RG equations under the initial conditions
that, at energies of the order of the cutoff, 
the response functions are finite. This gives us
the following equations. In the region where $c > c'$ where the
ferromagnetic response function dominates, neglecting the constant
term in (\ref{ferro}), and taking the renormalized couplings
as given by (\ref{intras}), (\ref{inters}), we have
\begin{equation}
\frac{\partial {\cal R}_{FM}}{\partial E_c}
\sim \frac{c}{2\pi^2 t}\;\;
\left[
\frac{U_0}{1 + U_0 \: c/(2 \pi^2 t) \:  
\log(\omega
       /E_c) }  \;+\;       
\frac{U_0}{1 + U_0 \: c'/(2 \pi^2 t) \:  
\log(\omega
       /E_c) }  \right] \;
\frac{1}{E_c}{\cal R}_{FM} \;\;,
\end{equation}
whose integration gives
\begin{equation}
{\cal R}_{FM}\sim {\cal R}_0 \; \frac{1}{1+ c/2\pi^2t \; U_0
\log \omega/E_c} 
\left( \frac{1}{1+ c'/2\pi^2t \; U_0
\log \omega/E_c}\right)^{c/c'} \;\;.
\end{equation}

The integration of the AFM  response function following the
same steps gives in the region $c'>c$ :

\begin{equation}
{\cal R}_{AFM}\sim {\cal R}_0 \; \frac{1}{1+ c'/2\pi^2t \; U_0
\log \omega/E_c} 
\left( \frac{1}{1+ c /2\pi^2t \; U_0
\log \omega/E_c}\right)^{c'/c} \;\;.
\end{equation}

The divergent flow of the superconducting response function is
given by 
\begin{equation}
{\cal R}_{SCd}\sim {\cal R}_0 \; 
\left(1+\frac{U_0 c}{2\pi^2 t}\log(\omega/E_c)\right)^{2\pi^2t/U_0c}
\frac{1}{\left(1+\frac{U_0 t}{2\pi^2 t}\log(\omega/E_c)\right)^{
2\pi^2 t\frac{c}{c'^2 U_0}}} \;\;.
\end{equation}

Putting the previous results all together, we arrive at the phase
diagram shown in fig. 12.

\vspace{0.5cm}

\begin{figure}
\epsfysize=15cm
\epsfbox{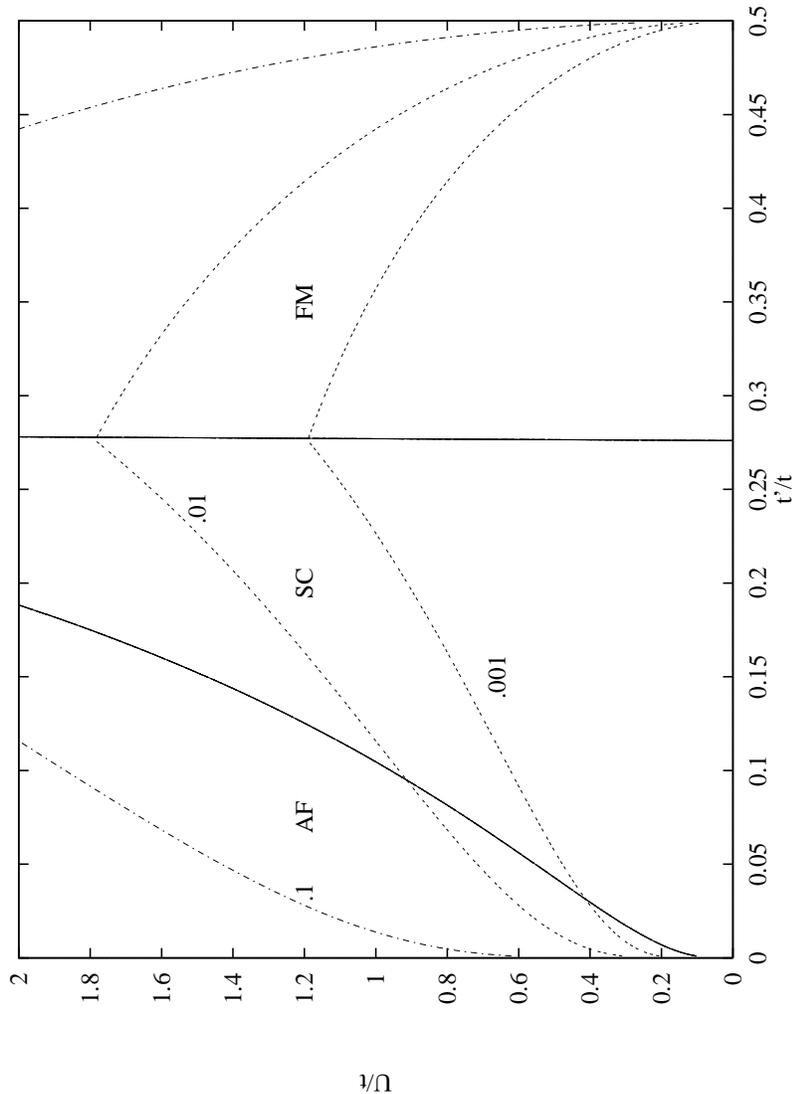}
\vspace{14 pt}
\caption{Phase diagram in the (t',U) plane. The dotted lines are contour 
lines corresponding to the critical frequencies shown in the figure.}
\label{phased}
\end{figure}
\vspace{0.5cm}

From inspection of Eq. (\ref{scd}), it is clear that divergent
correlations in the d-wave channel arise for $U_{intras} -
  U_{umk} < 0$. According to the above results, this only
happens for $c < c'$, that is, outside the region of the phase
diagram where $R_{FM} > R_{AFM}$. This confirms that a
ferromagnetic regime sets in for values of $t'$ above
the critical value $t_{c}' \approx 0.276 t $ at which $c = c'$. For
values below $t_{c}'$, there is a competition between $R_{AFM}$
and $R_{SCd}$, which requires the analysis of the respective
behaviors close to the critical frequency at which the response
functions diverge. As a general trend, the response function
$R_{SCd}$ dominates over $R_{AFM}$ in the regime of weak
interaction, the strength being measured with regard to both the
bare coupling constant and the value of the $c'$ parameter. The
reason for such behavior is that at weak interaction strength
the RG flow has a longer run to reach the critical frequency,
and at small frequencies the logarithmic density of states in
Eq. (\ref{scd}) makes $R_{SCd}$ to grow larger. The border where
the crossover between the antiferromagnetic and the superconducting
instability takes place is shown in the $t'-U$ phase diagram of
Fig. 10. At sufficiently large values of $U$ and
small values of $t'$, the leading instability of the system
turns out to be antiferromagnetism. This is in agreement with
weak coupling RG analyses applied to the Hubbard
model\cite{schulz,dzya,zanchi}.
Thus,  there exists a region of the phase diagram
where superconductivity is the leading instability. The
critical frequencies at which the instability takes place
are $\sim 10^{-2} E_c$. Taking values for $E_c$ of the order of
the conduction bandwidth in the cuprates, $\sim 1$eV, 
and $U \sim E_c$, we obtain critical temperatures $\sim 100$K.

The physical mechanism that induces an electronic pairing out of
purely repulsive interactions is the Kohn--Luttinger mechanism
\cite{kl}
according to which the competition between two different kinds
of --repulsive-- interactions,  can modulate the
net  interaction at large frquencies
giving rise to zones of attraction. This mechanism is known to
take place in D=3 for the usual Coulomb interaction with
spherical symmetry but there it becomes
effective  at very low temperatures and at very high
values of the angular momentum. In two dimensions
however and in a model like the one proposed here, due to the 
strong anisotropy of the Fermi line\cite{anis}, the
Kohn--Luttinger mechanism becomes much more efficient and can
drive superconductivity at frequencies accesible experimentally.

\section{Conclusions and open problems}

We start this section by mentioning the differences and
similarities between the present work and related works in the
literature. Concerning the papers that
apply RG techniques, we must establish a difference between
those appearing before and after the modern approach reviewed
in \cite{shankar}. In particular scaling methods were used
in \cite{schulz,zanchi} where they treat the whole
Fermi surface trying to arrive to a low--energy effective
action. What we do follows the approach described in \cite{pol}. 
We postulate a low--energy effective action valid only in the
proximity of the Fermi line, and study its stability under
renormalization. The same approach in the case of Fermi--liquid
systems yields surprisingly good results because the postulated
action turns out to be the attractive fixed point in all cases
--in the absence of BCS instability-- . In the present case it
is somehow unfortunate that the effective action is unstable in
the sense that all the interactions are marginally relevant and
we do not see a fixed point where it flows to.
This situation is very similar to what happens in systems of
coupled one--dimensional chains \cite{chains} where any
inter--chain coupling changes drastically the behavior of a
single chain. From our point of view, this behavior does not
invalidate the study presented in this paper as we believe that
the action (\ref{free}) will govern the behavior of any system
of the type discussed when the filling lies close to the Van
Hove singularities.

Concerning our previous work \cite{epl,npb}
the main difference with the present work is that we have
studied before spin--independent extended interactions --of the
type $V(r)$ described in the introduction -- and put the
emphasis in the possibility of
getting a pairing instability of electronic origin. In these cases
we obtained an attractive fixed point at the origin.
The present
work introduces the spin dependence and deals with
strictly local interactions what allows us to obtain the rich
phase diagram of fig. 12. It is the spin dependence of the
interactions what renders the action
unstable to all interactions.

As a summary of the results, we have shown that the purely repulsive 
Hubbard t--t' model at the Van Hove singularity exhibits a 
variety of instabilities at low
energy or temperature. In particular, we have seen
that it supports a superconducting
instability with d--wave symmetry coexisting in the same region
of parameter space with an antiferromagneting instability.
This feature has been recently found in
\cite{japo} by means of a mean--field computation. There is nowdays
a general agreement on the fact that d--wave
superconductivity is the strongest  experimental sign in hole doped
cuprates.

We remark that
this result is obtained within a RG approach that provides a
rigorous computational framework, with no other assumption than
the weakness of the bare interaction. The instabilities are led
by an unstable RG flow, and the singular behavior of the
response functions has been given a more solid 
foundation than the 
standard RPA computations. On the other hand, the renormalized
interactions remain during most part of the flow in the weak coupling
regime for $U < t$.

From a formal point of view, the second relevant conclusion
within our RG approach is the existence of a ferromagnetic
regime in the $t-t'$ model, above a certain value of the $t'$
parameter. This is consistent with the results obtained in
\cite{ferro} close to $t' = 0.5 t$. Though our results refer
to the weak coupling regime, they show that antiferromagnetic,
ferromagnetic and superconducting phases are all realized in the
$t-t'$ Hubbard model. The superconducting instability has
greater strength at the boundary with the antiferromagnetic
instability, as it also happens in other approaches to
high-$T_c$ superconductivity\cite{afm}.

In our case, the diagrams responsible for the apppearance of
superconductivity cannot be interpreted in terms of the exchange
of antiferromagnetic fluctuations, as in the
works mentioned earlier\cite{afm}. Those diagrams which contain
bubbles mediating an effective interaction between electron propagators
are cancelled, to all orders, by vertex corrections 
(see Fig. 3).
Superconductivity arises from the type of diagrams first studied
by Kohn and Luttinger\cite{kl}. The strong anisotropy of the Fermi
surface greatly enhances the Kohn-Luttinger mechanism, with
respect to its effect in an isotropic metal\cite{epl,npb,anis}.

The wide range for superconductivity is consistent with the
results from quantum Monte Carlo computations\cite{mc},
as well as with results obtained by exact diagonalization of
small clusters in the strong coupling regime\cite{nd}.
     
Our results support the idea that  d-wave superconductivity and 
antiferromagnetism arise from the same type of interactions.
Antiferromagnetism, however, does not favor the existence
of superconductivity, but competes with it in the same region
of parameter space. The renormalization group analysis
done in the present work does not allow to study the possible
existence of a quantum critical point at the end of the line
separating the antiferromagnetic and the superconducting phases
where a phase of higher symmetry has been postulated \cite{so5}.  
Similar physical processes seem to be
responsible for the appearance of anisotropic superconductivity
in systems of coupled repulsive 1D chains\cite{chains}.
The Fermi surface of a single chain is unable to give rise to
this type of superconductivity.  A soon as this limitation is
lifted,  superconductivity occupies a large fraction of the
phase diagram previously dominated by antiferromagnetic
fluctuations.

We now report on the open problems left in this work.

As a technical remark,
we have neglected self-energy corrections in the confidence based
on previous computations \cite{npb} that they will not change drastically
the Fermi-Liquid form. Our propagators are fermion propagators as we
expect the elementary excitations of the system to have a fermionic 
character. Here comes the subtle point of the nature of 
the free fixed point being an isolated point in the coupling constant
phase space (fig. 8). Unlike what happens in one dimensional 
systems where the quasiparticle pole is destroyed by the interaction, our 
interpretation of fig. 8 is that the Van Hove model has to be 
considered as an effective model  whose validity starts when the
Fermi surface of an otherwise Fermi liquid, sits close to the Van Hove 
singularity. Such a system will never show a Fermi liquid character
as the  instabilities described in the paper will take over.
This is confirmed by the experimental results. For overdoped samples,
the Fermi surface is closed and electron-like. In a non--interacting 
model, the Fermi surface would grow as more electrons are added until
it would reached the Van Hove points, acquiring a shape shown in
fig. 1. What is observed instead is that as more electrons are 
added, the system develops a gap near the Van Hove points \cite{dessau}.

The results reported in this work  can be extended to fillings away
from the van Hove singlarity, provided that the 
distance of the chemical potential to the singularity is
smaller than the energy scale at which the instability takes place.
The chemical potential tends to be pinned to the singularity
because of the non trivial RG flow of the chemical potential
itself\cite{epl,npb}. Hence, these calculations
can be applied to a finite range of fillings around that
appropiate to the singularity.

\vspace{1cm}
{\bf Acknowledgments} One of us (MAHV) thanks the Institute of 
Advanced Study of Princeton for its hospitality during the summer of
1997 when the manuscript was completed.

\end{document}